\documentstyle[epsf]{mn}

\input{psfig}  %ps figures

\newif\ifAMStwofonts
\newcommand{\be}{\begin{equation}}
\newcommand{\ee}{\end{equation}}
\newcommand{\ba}{\begin{eqnarray}}
\newcommand{\ea}{\end{eqnarray}}
\newcommand{\brr}{\begin{array}}
\newcommand{\err}{\end{array}}
\newcommand{\bc}{\begin{center}}
\newcommand{\ec}{\end{center}}
\newcommand{\hMpc}{\mbox{$h^{-1}{\rmn{Mpc}}~$}}

\newcommand{\lum}{\,{\rm erg\,s^{-1}}}

\newcommand{\mincir}{\raise
  -2.truept\hbox{\rlap{\hbox{$\sim$}}\raise5.truept \hbox{$<$}\ }}
\newcommand{\magcir}{\raise
  -2.truept\hbox{\rlap{\hbox{$\sim$}}\raise5.truept \hbox{$>$}\ }}
\newcommand{\siml}{\raise
  -2.truept\hbox{\rlap{\hbox{$\sim$}}\raise5.truept \hbox{$<$}\ }}
\newcommand{\simg}{\raise
  -2.truept\hbox{\rlap{\hbox{$\sim$}}\raise5.truept \hbox{$>$}\ }}

\ifoldfss
\newcommand{\rmn}[1] {{\rm #1}}

\ifCUPmtlplainloaded \else
\NewTextAlphabet{textbfit} {cmbxti10} {}
\NewTextAlphabet{textbfss} {cmssbx10} {}
\NewMathAlphabet{mathbfit} {cmbxti10} {} % for math mode
\NewMathAlphabet{mathbfss} {cmssbx10} {} %  "   "    "
\fi
\ifAMStwofonts
\ifCUPmtlplainloaded \else
\NewSymbolFont{upmath} {eurm10}
\NewSymbolFont{AMSa} {msam10}
\NewMathSymbol{\upi}     {0}{upmath}{19}
\NewMathSymbol{\umu}     {0}{upmath}{16}
\NewMathSymbol{\upartial}{0}{upmath}{40}
\NewMathSymbol{\leqslant}{3}{AMSa}{36}
\NewMathSymbol{\geqslant}{3}{AMSa}{3E}

 \let\le=\leqslant
 \let\ge=\geqslant
\fi
\fi
\fi % End of OFSS

\ifnfssone
\newmathalphabet{\mathit}
\addtoversion{normal}{\mathit}{cmr}{m}{it}
\addtoversion{bold}{\mathit}{cmr}{bx}{it}
\newcommand{\rmn}[1] {\mathrm{#1}}

\newmathalphabet{\mathbfit} % math mode version of \textbfit{..}
\addtoversion{normal}{\mathbfit}{cmr}{bx}{it}
\addtoversion{bold}{\mathbfit}{cmr}{bx}{it}
\newmathalphabet{\mathbfss} % math mode version of \textbfss{..}
\addtoversion{normal}{\mathbfss}{cmss}{bx}{n}
\addtoversion{bold}{\mathbfss}{cmss}{bx}{n}
\ifAMStwofonts
\ifCUPmtlplainloaded \else
\UseAMStwoboldmath
\makeatletter
\new@mathgroup\upmath@group
\define@mathgroup\mv@normal\upmath@group{eur}{m}{n}
\define@mathgroup\mv@bold\upmath@group{eur}{b}{n}
\edef\UPM{\hexnumber\upmath@group}
\new@mathgroup\amsa@group
\define@mathgroup\mv@normal\amsa@group{msa}{m}{n}
\define@mathgroup\mv@bold\amsa@group{msa}{m}{n}
\edef\AMSa{\hexnumber\amsa@group}
\makeatother
\mathchardef\upi="0\UPM19
\mathchardef\umu="0\UPM16
\mathchardef\upartial="0\UPM40
\mathchardef\leqslant="3\AMSa36
\mathchardef\geqslant="3\AMSa3E

 \let\le=\leqslant
 \let\ge=\geqslant
\fi
\fi
\fi % End of NFSS release 1

\ifnfsstwo
\newcommand{\rmn}[1] {\mathrm{#1}}

\DeclareMathAlphabet{\mathbfit}{OT1}{cmr}{bx}{it}
\SetMathAlphabet\mathbfit{bold}{OT1}{cmr}{bx}{it}
\DeclareMathAlphabet{\mathbfss}{OT1}{cmss}{bx}{n}
\SetMathAlphabet\mathbfss{bold}{OT1}{cmss}{bx}{n}
\ifAMStwofonts
\ifCUPmtlplainloaded \else
\DeclareSymbolFont{UPM}{U}{eur}{m}{n}
\SetSymbolFont{UPM}{bold}{U}{eur}{b}{n}
\DeclareSymbolFont{AMSa}{U}{msa}{m}{n}
\DeclareMathSymbol{\upi}{0}{UPM}{"19}
\DeclareMathSymbol{\umu}{0}{UPM}{"16}
\DeclareMathSymbol{\upartial}{0}{UPM}{"40}
\DeclareMathSymbol{\leqslant}{3}{AMSa}{"36}
\DeclareMathSymbol{\geqslant}{3}{AMSa}{"3E}

 \let\le=\leqslant
 \let\ge=\geqslant
\fi
\fi
\fi % End of NFSS release 2

\ifCUPmtlplainloaded \else
\ifAMStwofonts \else % If no AMS fonts
\def\upi{\pi}
\def\umu{\mu}
\def\upartial{\partial}
\fi
\fi

\title[Cosmological Constraints from the $\xi(r)$ of the XBACs]
{Cosmological Constraints from the Clustering
Properties of the $X$--Ray Brightest Abell--type Cluster
  Sample} 
\author[S.Borgani, M. Plionis, V. Kolokotronis]
{S. Borgani$^1$\footnote{Email: Stefano.Borgani@pg.infn.it},
M. Plionis$^2$\footnote{Email: plionis@sapfo.astro.noa.gr}
and V. Kolokotronis$^{2,3}$\footnote{Email: vk@sapfo.astro.noa.gr}\\
$^1$ INFN, Sezione di Perugia, c/o Dipartimento di Fisica
dell'Universit\`a, via A. Pascoli, I-06123 Perugia, Italy\\
$^2$ National Observatory of Athens, Lofos Nimfon, Thesio, 18110
Athens, Greece\\
$^3$ Astronomy Unit, School of Mathematical Sciences, Queen Mary \&
Westfield College, Mile End Road, London E1, 4NS, UK\\
}
\pagerange{\pageref{firstpage}--\pageref{lastpage}}
\pubyear{1998}

\begin{document}
\label{firstpage}
\maketitle

\begin{abstract}
  We present an analysis of the 2--point correlation function,
  $\xi(r)$, of the $X$--ray Brightest Abell--type Cluster sample (XBACs;
  Ebeling et al. 1996) and of the cosmological constraints that it
  provides.  If $\xi(r)$ is modelled as a power--law,
  $\xi(r)=(r_0/r)^\gamma$, we find $r_0 \simeq 26.0\pm 4.5$ \hMpc and
  $\gamma\simeq 2.0\pm 0.4$, with errors corresponding to 2$\sigma$
  uncertainties for one significant fitting parameter. As
  a general feature, $\xi(r)$ is found to remain positive up to
  $r\simeq 50$--55 \hMpc, after which it declines and crosses zero.
  Only a marginal increase of the correlation amplitude is found as
  the flux limit is increased from $5\times 10^{-12}$ erg s$^{-1}$
  cm$^{-2}$ to $12\times 10^{-12}$ erg s$^{-1}$ cm$^{-2}$, thus
  indicating a weak dependence of the correlation amplitude on the
  cluster $X$--ray luminosity. Furthermore, we present a method to predict
  correlation functions for flux--limited $X$--ray cluster samples
  from cosmological models. The method is based on the analytical
  recipe by Mo \& White (1996) and on an empirical approach to convert
  cluster fluxes into masses. We use a maximum--likelihood method to
  place constraints on the model parameter space from the XBACs
  $\xi(r)$. For scale--free primordial spectra, we find that the shape
  parameter of the power spectrum is determined to lie in the
  $2\sigma$ range $0.05\mincir \Gamma \mincir 0.20$. As for the
  amplitude of the power--spectrum, we find $\sigma_8\simeq 0.4$--0.8
  for $\Omega_0=1$ and $\sigma_8\simeq 0.8$--2.0 for $\Omega_0=0.3$.
  The latter result is in complete agreement with, although less
  constraining than, results based on the local cluster abundance.
\end{abstract}


\section{Introduction}
Galaxy clusters have been recognised since a long time as extremely
useful tracers of the large--scale structure of the Universe. Being
the largest virialized cosmic structures, they can be detected rather
unambiguously up to very large distances. Furthermore, their
distribution traces scales which have not yet undergone the
non--linear phase of gravitational clustering, therefore simplifying
their connection to the initial conditions of cosmic structure
formation.

So far, most of the studies of the cluster
distribution have relied on optical samples like the Abell/ACO one (Abell
1958; Abell Corwin \& Olowin 1989), the APM sample (Dalton et al.
1994) and the EDCC sample (Collins et al. 1995). A general result
from such analyses is that the two--point cluster correlation function
is well approximated by a power--law, $\xi(r)=(r_0/r)^\gamma$, with
$\gamma \simeq 1.8$--2. The measured correlation length $r_0$,
range in the interval $r_0\simeq 15$--25 \hMpc,
depending on the analysed sample and/or on the cluster richness (see,
e.g., Bahcall \& West 1992, Croft et al. 1997; and references
therein).

A serious problem arising in optical cluster compilations is the
spurious enhancement of the cluster richness in high density
environments which can also cause inherently poor clusters or groups
to appear rich enough to be included in the sample (e.g. van Haarlem,
Frenk \& White 1997).  This led several authors to cast doubts about
the reliability of the clustering analysis of Abell/ACO clusters,
whose correlation function could be artificially enhanced by such
projection effects (cf.  Sutherland 1988; Dekel et al. 1989; Peacock
\& West 1992; see, however, Jing, Plionis \& Valdarnini 1992 for an
alternative view). Although samples like the APM and the EDCC have
been designed with the purpose of minimising projection effects, they
could still be present at some level (e.g., Collins et al. 1995).

This calls for the need to confirm results based on optical samples by
studying other cluster compilations which are free of the above
mentioned biases. $X$--ray selected cluster samples are ideally suited
for this purpose. Indeed, since the $X$--ray emissivity of the
intra--cluster medium (ICM) is proportional to the square of the gas
density, cluster emission is strongly peaked at the centre, so as to
make the impact of projection effects almost negligible. Important
analyses of the large-scale distribution of relatively small $X$--ray
cluster samples have already been performed in recent years (cf. Romer
et al. 1994; Nichol, Briel \& Henry 1994, Guzzo et al. 1995).

More recently, the $X$--Ray Brightest Abell Cluster sample (XBACs,
hereafter; Ebeling et al. 1996) has been compiled from
cross--correlating the {\sl ROSAT} all--sky $X$--ray survey with the
Abell/ACO cluster sample (for further details see their Section 4).
Although it is not purely $X$--ray selected, the inclusion of only
those Abell/ACO clusters that have an extended and significant $X$-ray
emission suppress the biases affecting the purely optical sample (cf.
Ebeling et al. 1996, Plionis \& Kolokotronis 1998; PK98, hereafter).
The XBACs provides for the first time a whole sky, flux--limited
sample of $X$--ray galaxy clusters. This sample has been analysed by
PK98 with the aim of investigating the local acceleration field and by
Abadi, Lambas \& Muriel (1998) to study the cluster 2--point
correlation function (see also Kolokotronis 1998).

In this paper, we present a new analysis of the correlation function
for the XBACs cluster distribution, with the specific aim of investigating
its implication on cosmological models for large--scale structure
formation. In this context, a great advantage in using galaxy
clusters, instead of galaxies, as tracers of the large-scale structure
lies in their less ambiguous connection with the underlying dark
matter (DM) density fluctuations. As for $X$--ray clusters, the problem of
understanding their biasing with respect to the DM distribution 
requires a suitable description of the
connection between the $X$--ray luminosity and the mass of the
virialized DM halo hosting the emitting gas. This problem, although
not of trivial solution, is much more affordable, both from a
numerical and an analytical point of view, than an accurate
understanding of galaxy formation and evolutionary processes.

In what follows, we will use the analytical approach by Mo \& White
(1996) to describe the correlation function for the distribution of
clusters, which are identified as virialized halos of a given mass.
This recipe in itself is not enough to compare model predictions with
XBACs data, since the XBACs cluster selection is clearly not based on
a mass threshold criterion.  For this reason, we will resort to a
phenomelogical approach to connect cluster masses to observed $X$--ray
fluxes (cf. also Borgani et al. 1998). Although our analysis is
entirely based on a sample of local clusters, all the formalism that
we will present is developed to treat the cluster distribution at a
generic redshift. Therefore, our analysis method is amenable to be
directly applied to obtain cosmological constraints from future deep
surveys of galaxy clusters.

The structure of the paper can be summarised as follows. In Section 2
we provide a short description of the XBACs sample and of the results
of the $\xi(r)$ analysis. We describe in Section 3 how to compute
cluster correlations for an $X$--ray flux--limited sample from
cosmological models, which are characterised by a given
power--spectrum of density fluctuations and by the geometry of the
Friedmann background. The method to compare model predictions to real
data analysis is discussed in Section 4. In Section 5 we briefly
discuss such results and draw the main conclusions.

\section{Correlation analysis of XBACS}
\subsection{The XBACs sample}
The XBACs sample consists of the X--ray brightest Abell/ACO clusters
that have been detected in the {\small ROSAT} all sky survey (RASS;
Tr\"umper 1990; Voges 1992) for fluxes above $S_{\rm lim}=5 \times
10^{-12}\,$erg s$^{-1}\,$cm$^{-2}$ ([0.1--2.4] keV band) with
redshifts limited by $z\le 0.2$. The sample contains 253 clusters and
thus it is the largest X--ray flux--limited cluster sample to date
(though not entirely X--ray selected).  The X--ray fluxes measured
initially by the Standard Analysis Software System (SASS) point source
detection algorithm are superseded by Voronoi Tessellation Percolation
(VTP) measurements, which better account for the extended nature of the
emission. In addition, the difficulty of the SASS algorithm to
actually detect nearby X--ray emission has been mostly corrected by
running VTP on the {\small RASS} fields centred on the optical
positions of all nearby Abell/ACO clusters ($z\le\,$0.05) irrespective
of whether or not they are detected by SASS. Ebeling et al. (1996)
estimate, after a careful analysis of possible selection effects and
biases that the overall completeness of this X--ray sample is 
$\magcir 80\%$ at the above flux limit.

An extended discussion of the possible XBACs systematic effects 
is presented in PK98. Here we present only 
a brief summary.

The known volume incompleteness effect, as a function of distance, for
the richness class R=0 Abell/ACO clusters, is not present in the XBACs
sample (Ebeling et al. 1996). Indeed, thanks to its flux--limited
property, the sample contains at large distances the inherently
brighter and thus richer Abell/ACO clusters, for which there is no
volume incompleteness.  Furthermore the significant distance dependent
density variations, which are most probably due to the higher
sensitivity of the ACO IIIa-J emulsion plates, between the northern
Abell and southern ACO parts of the combined optical cluster sample
(cf. Batuski et al. 1989; Scaramella et al. 1991; Plionis \&
Valdarnini 1991) has been corrected for in the XBAC sample (see
details in PK98).

The main remaining biases that need to be quantified for the correlation 
analysis are the redshift selection function, which depends on the XBACs
luminosity function and flux limit, and the Galactic absorption
latitude selection function, which has been found to be consistent with
a cosecant law of the form
\begin{equation}\label{eq:obs}
P(|b|) = \mbox{dex} \; \left[ \alpha \left(1 - \csc |b| \right) \right]
\end{equation}
with $\alpha \approx 0.3$ for the Abell sample (Bahcall \& Soneira 1983; 
Postman et al. 1989) and $\alpha \approx 0.2$ for the ACO sample 
(Batuski et al. 1989). In the following we will restrict our analysis
to clusters having $|b|\ge 30^\circ$.

\begin{figure}
%\vspace{5cm}
\centerline{
\psfig{figure=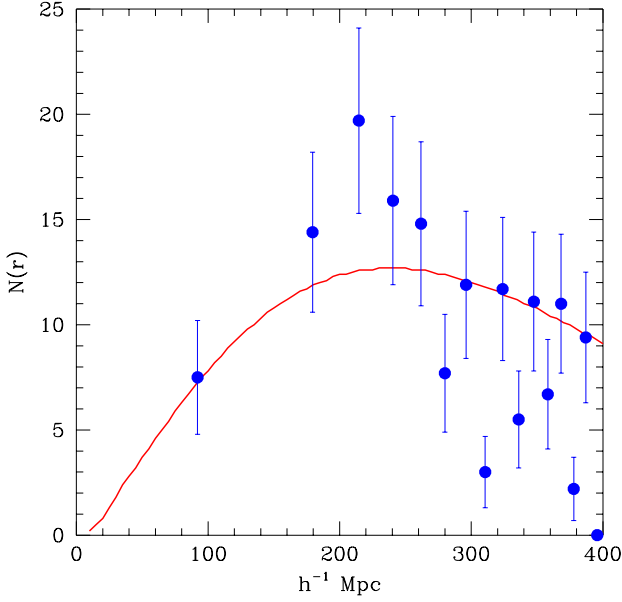,height=8.5cm}
}
%\vspace{-6.5truecm}
\caption{The XBACs radial distribution function and its Poisson uncertainty 
(filled circles). The predicted distribution based on the XBACs 
luminosity function estimated by PK98 is shown as a continuous line.}
\label{fi:Nr}
\end{figure}

The XBACs redshift distribution, in the form of an $N(r)$ plot,
corrected for Galactic absorption, can be seen in Figure \ref{fi:Nr}.
We have also plotted as a continuous line the predicted radial
distribution function based on the XBACs luminosity function (see
references in PK98) and the $5\times 10^{-12}$ erg s$^{-1}$ cm$^{-2}$
flux limit.
Note that redshifts are converted into luminosity distances according to
\be
d_L(z)\,=\,{c\over H_0}\,r(z)\,(1+z)
\label{eq:dis}
\ee
where $H_0=100\,h$ km s${-1}$ Mpc$^{-1}$ is the Hubble constant and
\ba
r(z) & = & \int_0^z dz \,E^{-1}(z) ~~~;~~~\Omega_\Lambda=1-\Omega_0 \nonumber \\
r(z) & = & {2\left[\Omega_0z+(2-\Omega_0)\,(1-\sqrt{1+\Omega_0z})\right]\over
  \Omega_0^2(1+z) } ~;~ \Omega_\Lambda=0\,.
\label{eq:rz}
\ea 
Here $E(z)=[(1+z)^3\Omega_0+\Omega_\Lambda]^{1/2}$ for a flat Universe 
(e.g., Peebles 1993).
In what follows, results and comparisons with models will be
presented for two values of the density parameter; $\Omega_0=1$ and 0.3,
with and without a cosmological constant term ($\Omega_\Lambda=\Lambda/
3H_0^2$) to restore spatial flatness. 

\subsection{The $\xi(r)$ estimate}
We estimate the 2-point correlation function for XBACs using the 
following estimator:
\begin{equation}
1 + \xi(r) = 2\,\frac{N_{cc}}{\langle N_{cr} \rangle} 
\end{equation}
where $N_{cc}$ is the number of cluster pairs in the interval
$[r-\delta r, r+\delta r]$ and $\langle N_{cr} \rangle$ is the
average, over 10,000 random simulations with the same boundaries,
redshift and galactic latitude selection functions,
cluster-random pairs in the same separation interval.  We have
evaluated $\xi(r)$ in logarithmic intervals with constant logarithmic
amplitude $\Delta \sim 0.1$. We estimate the variance of $\xi(r)$ by using
the analytical approximation to the bootstrap errors (Mo, Jing \& 
B\"orner 1992) which has been shown to reproduce fairly accurately the actual
bootstrap variance:
\begin{equation}\label{eq:mo}
\sigma_\xi^{2} \simeq 3 \times \sigma_{qP}^{2}=3\times {1+\xi(r)\over
N_{cc}(r)} \,,
\end{equation}
where $\sigma^{2}_{qP}$ is the quasi--Poisson variance.

\begin{table}
\centering
\caption[] {Results for the three samples. Column 2: flux limits
  (units of $10^{-12}\,$erg s$^{-1}\,$cm$^{-2}$); Column 3: number of
clusters; Column 4: correlation length and 2$\sigma$ uncertainty
(units of \hMpc); Column 5: slope of $\xi(r)$ and 2$\sigma$
uncertainty.}
\tabcolsep 5pt
\begin{tabular} {cccccc}
  & $S_{lim}$ & No. of clusters & $r_0$ & $\gamma$ & $\chi^2_{min}$  \\
 ~ \\
\noalign{\smallskip}
(a) & 5 & 203 & $26.0^{+4.1}_{-4.7}$ &  $1.98^{+0.35}_{-0.53}$ & 3.9 \\ ~\\
\noalign{\smallskip}
(b) & 8 & 112 & $25.9^{+5.2}_{-6.6}$ &  $2.02^{+0.43}_{-0.76}$ & 3.6 \\ ~\\
\noalign{\smallskip}
(c) & 12 & 67 & $27.5^{+8.7}_{-12.8}$ & $1.94^{+0.85}_{-1.52}$ & 5.8
\end{tabular}
\label{t:sam}
\end{table}

We apply the correlation analysis to the whole sample ($z\mincir
0.2$), with $S_{\rm lim}=5 \times 10^{-12}\,$erg s$^{-1}\,$cm$^{-2}$
[sample {\sl (a)}] and to two other subsamples with $S_{\rm lim}=8
\times 10^{-12}\,$erg s$^{-1}\,$cm$^{-2}$ and $S_{\rm lim}=12 \times
10^{-12}\,$erg s$^{-1}\,$cm$^{-2}$ [samples {\sl (b)} and {\sl (c)},
respectively].  The total number of clusters in each sample is
reported in Table \ref{t:sam}.

\begin{figure*}
%\vspace{5cm}
\centerline{
%\mbox{\psfig{figure=xi_all.ps,height=15cm,bbllx=18pt,bblly=144pt,bburx=592pt,bbury=718pt}}
\psfig{figure=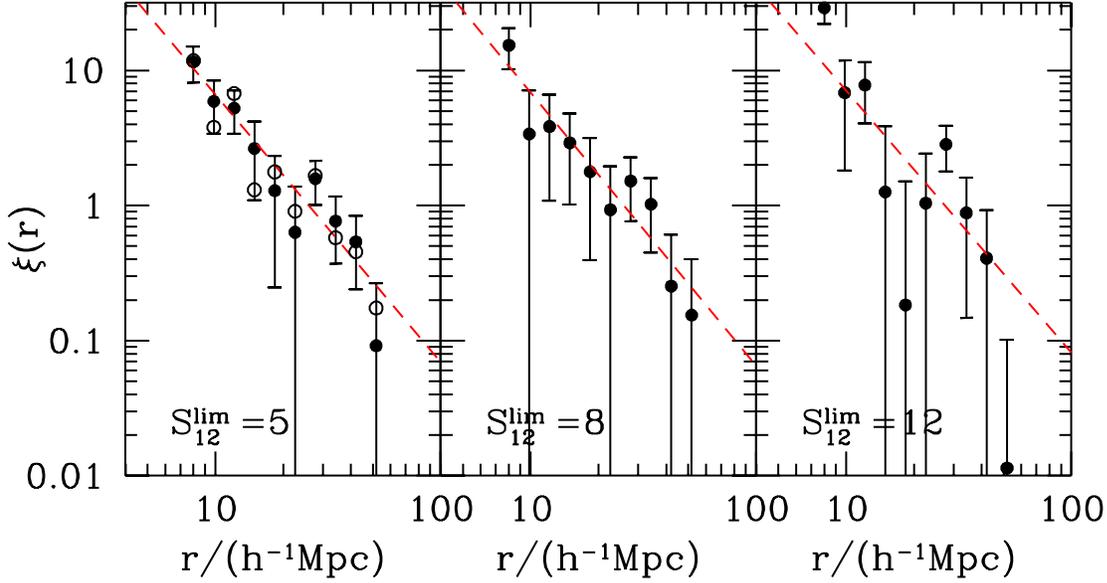,height=15cm}
}
\vspace{-6.5truecm}
\caption{The 2--point correlation function for samples {\sl (a), (b)}
  and {\sl (c)}, by assuming $\Omega_0=1$ to convert redshifts into
  distances (cf. eqs.(\ref{eq:dis}) and (\ref{eq:rz}). Errorbars
  correspond to 1$\sigma$ bootstrap uncertainties.  Open circles for
  the {\sl (a)} sample corresponds to the open $\Omega_0=0.3$ case.
  The dashed lines corresponds to the best fitting power--law
  expressions, $\xi(r)=(r_0/r)^\gamma$, whose parameters are reported
  in Table \ref{t:sam}.}
\label{fi:xi}
\end{figure*}

The resulting $\xi(r)$, using $\Omega_0=1$ in eq.(\ref{eq:rz}), for the 
three samples are plotted in Figure \ref{fi:xi}. Only for sample 
{\sl (a)} we show the $\xi(r)$ by using $\Omega_0=0.3$. 
Due to the limited depth of XBACs no significant differences are found 
between these two cases.
For all the three samples $\xi(r)$ is positive up to $r\simeq 50$ \hMpc, 
after which it declines and crosses zero at about 50 - 55 \hMpc. 
This result agrees with that
found for optical Abell/ACO cluster samples (e.g., Klypin \& Rhee
1994) and confirms that the cluster distribution requires a
substantial amount of large--scale coherence of cosmic density
fluctuations.

\begin{figure}
%\vspace{5cm}
\centering
%\mbox{\psfig{figure=xi_pow.ps,height=15cm,bbllx=18pt,bblly=144pt,bburx=592pt,bbury=718pt}}
\psfig{figure=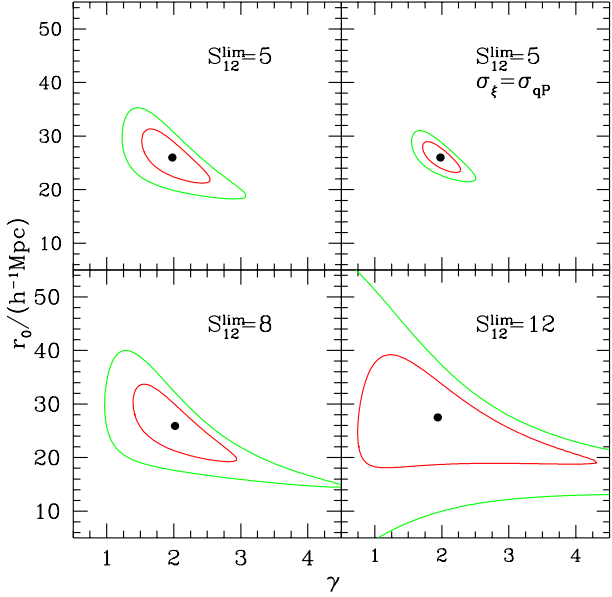,height=8.5cm}
\caption{Iso--$\Delta\chi^2$ contours on the $r_0$--$\gamma$ parameter
  space from the correlation analysis of samples {\sl (a), (b)} and
  {\sl (c)} (upper left, lower left and lower right panels,
  respectively). The upper right panel show the results for sample
  {\sl (a)} if quasi--Poissonian errors for $\xi(r)$ were instead
  assumed. Contours correspond to $1\sigma$, $2\sigma$ and $3\sigma$
  confidence levels for 2 significant fitting parameters.}
\label{fi:fitpow_conf}
\end{figure}

\begin{figure}
%\vspace{5cm}
\centering
%\mbox{\psfig{figure=xi_pow.ps,height=15cm,bbllx=18pt,bblly=144pt,bburx=592pt,bbury=718pt}}
\psfig{figure=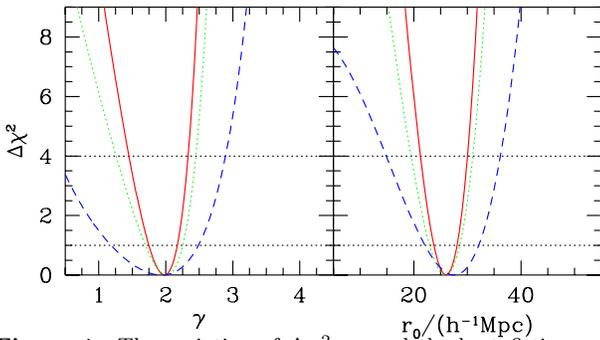,height=8.5cm}
\vspace{-4.truecm}
\caption{
The variation of
$\Delta \chi^2$ around the best--fitting value of each parameter, after
marginalising with respect to the other parameter. Continuous, dotted
and dashed curves are for {\sl (a), (b)} and {\sl (c)} samples,
respectively. The two horizontal lines and the upper limit of each
panel represent the $1\sigma$, 2$\sigma$ and 3$\sigma$ confidence
levels for one significant parameter.}
\label{fi:fitpow_marg}
\end{figure}

The dashed lines in the plots represent the best--fitting power law
model, $\xi(r)=(r_0/r)^\gamma$, which is determined through a
$\chi^2$--minimisation ($\chi^2_{min}$ hereafter) procedure, assuming
a Gaussian distribution for $\sigma_\xi$.  The fit for the {\sl (c)}
sample has been performed only including bins with $r> 10$ \hMpc, so
as to exclude the anomalous high--correlation signal coming from the
smallest scales.  On a more quantitative ground, Figure
\ref{fi:fitpow_conf} shows the iso--$\Delta\chi^2$ contours ($\Delta
\chi^2=\chi^2-\chi^2_{min}$, with $\chi^2_{min}$ being the absolute
minimum value of $\chi^2$) in the $r_0$--$\gamma$ plane.  Strictly
speaking, $\xi(r)$ estimates in different bins are not independent of
each other, since each cluster contributes to pairs at different
separations.  The contours correspond to $1\sigma$ and $2\sigma$
uncertainties for two significant parameters and correspond to
$\Delta\chi^2=2.30$ and 6.17, respectively. Figure
\ref{fi:fitpow_marg} shows the variation of $\Delta\chi^2$ around the
best--fitting value of each of the two parameters, once we marginalise
with respect to the other parameter. The best fitting values for $r_0$
and $\gamma$, along with the $2\sigma$ uncertainty for one significant
fitting parameter are also reported in Table \ref{t:sam}. It turns out
that increasing $S_{lim}$ the correlation length only marginally
increases. As long as a correlation exists between cluster $X$--ray
luminosity and richness, this result implies only a mild dependence of
the clustering strength on the cluster richness (see also Croft et al.
1997).

Results for sample {\sl (a)} can be compared with those obtained
by Abadi et al. (1998) for the same sample. They find a somewhat lower
correlation amplitude, with $r_0\simeq 21$ \hMpc and a similar $\gamma$ value
but with much smaller uncertainties (cf. their Fig. 4). We are not sure 
why their estimate of $r_0$ is lower than ours. However,
the rather large difference in the uncertainties is entirely due to their
assumption of Poissonian errors for $\xi(r)$. We verified that
repeating our analysis assuming quasi--Poisson errors (cf. the
upper right panel in Fig. \ref{fi:fitpow_conf}), the contours in
the $r_0$--$\gamma$ plane shrinks into a much smaller size, comparable to
that reported by Abadi et al. (1998). 

\section{Predicting correlations for flux--limited samples}
In this section we introduce the formalism to compute model $\xi(r)$
for a flux--limited cluster survey. After briefly introducing the analytical
method by Mo \& White (1996) to estimate correlations for virialized
halos of a given mass, we present an empirical procedure to convert
this mass into an $X$--ray flux in the appropriate energy band. A
similar approach has been also applied by Moscardini et al. (1999, in
preparation) to predict the 2--point correlation functions expected in
different $X$--ray flux--limited cluster samples. We
point out that, although our method is applied here to the analysis of
the local XBACs clusters, it is presented in a general form so as
to be directly applicable to any sample of both local and distant
$X$--ray selected clusters. Therefore, it is also well
suited to study the clustering evolution using future deep cluster
surveys.

\subsection{The analytical recipe for cluster correlations}
Our estimates of model cluster correlations is based on the approach
originally proposed by Mo \& White (1996; MW hereafter).  In this
approach, the correlation function at redshift $z$ for clusters
(identified as virialized halos) of mass $M$, $\xi_{cl}(r,z,M)$, is
connected to the dark matter correlation function at the same
redshift, $\xi_{DM}(r,z)$, according to 
\be
\xi_{cl}(r,z,M)\,=\,b^2(z,M)\,\xi_{DM}(r,z)
\label{eq:xih}
\ee
where
\be
\xi_{DM}(r,z)\,=\,\left[{D(z)\over D(0)} \right]^2{1\over
    2\pi^2}\int_0^\infty dk\,P(k)\,k^2\,{\sin kr\over kr}\,.
\label{eq:xidm}
\ee 
In the above expression $D(z)$ is the linear growth factor for
density fluctuations at redshift $z$ (e.g., Peebles 1993).

The biasing factor, $b(M,z)$, appearing in eq.(\ref{eq:xih}) is given by
\be
b(M,z)\,=\,1+{[\delta_c(z)/\sigma(R_M,z)]^2-1\over \delta_c(z)}
\label{eq:bias}
\ee
(e.g, MW; Matarrese et
al. 1997; Baugh et al. 1998). Here $\delta_c(z)$ is the critical
linear overdensity for spherical collapse.  For a critical--density
Universe $\delta_c=1.686$, independent of $z$, with a weak
dependence of the geometry of the Friedmann background
(e.g., Eke, Cole \& Frenk 1996). Furthermore,
\be
\sigma^2(R_M,z)\,=\,\left[{D(z)\over D(0)}\right]^2
{1\over 2\pi^2}\,\int_0^\infty dk\,k^2\,P(k)\,W^2(kR_M)\,.
\label{eq:sigm}
\ee
is the fluctuation variance at mass $M$ and redshift $z$,
with $W(x)$ the Fourier representation of the window function,
which describes the shape of the volume from where the collapsing
object accreates matter. The comoving fluctuation size $R_M$ represents 
the Lagrangian cluster radius. It is
connected to the mass scale $M$ and to the present day average matter
density $\bar \rho$ according to
\be
R_M\,=\,\left({3M\over 4\pi \bar\rho}\right)^{1/3}
\label{eq:lr}
\ee
for the top--hat window, $W(x)=3(\sin x- x\cos x)/x^3$, that we adopt
in the following. 

Since in practical cases one is interested in the correlation of halos
with mass {\em above} a given limit $M$, the biasing factor in
eq.(\ref{eq:bias}) should be replaced by an {\em effective} bias,
$b_{eff}$, whose value is obtained by averaging $b$ over the mass
distribution of the virialized halos:
\be
b_{eff}(M,z)\,=\,{\int_M^\infty dM' \,b(M',z)\,n(M',z) \over
  \int_M^\infty dM' \,n(M',z)}\,.
\label{eq:beff}
\ee
In the above expression the mass distribution $n(M,z)$ is given by the
Press \& Schechter (1974) expression
\ba
&& n(M,z)\,dM\,= \,\sqrt{2\over \pi}\, {\bar \rho \over M^2}\,
{\delta_c(z)\over \sigma(M,z)}\, \left|{d\log \sigma(M,z)\over d\log
    M}\right|\, \nonumber \\
&&  \exp\left(-{\delta_c^2(z)\over
    2\sigma^2(M,z)}\right)\,dM\,. 
\label{eq:ps}
\ea 
The power--spectrum $P(k)$ of density fluctuations can be
expressed as $P(k)=Ak^nT^2(k)$, where $n$ is the primordial spectral
index. A Harrison-Zel'dovich primordial spectrum, with $n=1$, will be
assumed in the following. As for the transfer function, $T(k)$, we
take the expression
\ba
&&T(q)\,=\,{{\rm ln}(1+2.34 q)\over 2.34 q}\times \nonumber \\
&& \left[1+3.89q+(16.1q)^2+(5.46q)^3+(6.71q)^4\right]^{-1/4}\,.
\label{eq:tk}
\ea 
Here $q=k/h\Gamma$, $\Gamma$ being the shape parameter. For the
class of CDM models, it is $\Gamma\simeq \Omega_0h$ (Bardeen et al. 1986),
while in general $\Gamma$ can be viewed as a free parameter, to be
fitted against observational constraints. For instance, the power--spectrum
of APM galaxies provide $\Gamma = 0.23\pm 0.04$ (e.g., Peacock \&
Dodds 1996). As for the amplitude of $P(k)$, it is customary to
express it in terms of $\sigma_8$, the r.m.s. fluctuation amplitude
within a top--hat sphere of 8 \hMpc radius. 

Therefore, each model for large--scale structure formation will be
characterised by three parameters, namely $\Omega_0,\sigma_8$ and
$\Gamma$. In the following, we will present results on the
$\Gamma$--$\sigma_8$ plane, fixing the density parameter to either 
$\Omega_0=0.3$ or $\Omega_0=1$.

The reliability of the MW approach to describe the clustering of
cluster--sized halos has been already tested by Mo, Jing \& White
(1996; cf. also Governato et al. 1998).  Jing (1998) recently showed
that a correction to eq.(\ref{eq:bias}) is required only for halos of
mass much smaller than $M^*$ (defined as the mass for which
$\delta_c/\sigma_M=1$), a regime which is not relevant for clusters in
plausible cosmological models.  Colberg et al. (1998) found from the
analysis of their Hubble Volume Simulations that the MW method
overpredicts the cluster correlation length $r_0$ by $\sim 20\%$.  In
the following analysis we will compute the biasing factor according to
eq.(\ref{eq:beff}) by assuming for $\delta_c$ the canonical value of
the spherical top--hat collapse model. We just point out that, since
$b_{eff}$ depends on $\delta_c$ and $\sigma_M$ only through their
ratio, a change in the choice of $\delta_c$ turns into a proportional
change in the resulting values of $\sigma_8$.

Finally, since the real data are analysed in redshift space, 
we introduce the effect of $z$--distortion in the MW expression for
$\xi(r)$. Redshift--space correlations are amplified by
the usual factor $K(\beta)=1+2\beta/3+\beta^2/5$ (Kaiser 1987), where
$\beta\simeq \Omega_0^{0.6}/b_{eff}$. However, for the cluster case
where $\beta \sim 0.20$, the effect is rather small since it increases 
$\xi(r)$ by only $\sim 15\%$.

\subsection{Converting fluxes into masses}
Cluster masses are the input quantities for the MW analytical approach
that we have just described. Since the observable quantity is the cluster
luminosity rather than its mass we should provide  a suitable method 
to convert mass into fluxes (or vice-versa).
 
As a first step we convert masses into $X$--ray temperatures.
According to the spherical collapse model and under the assumption of virial
equilibrium, the mass--temperature relation can be written as
\ba
k_BT & = & {1.38\over \beta_T}\,\left({M\over
10^{15}h^{-1}M_\odot}\right)^{2/3}\nonumber \\
& \times &\left[\Omega_0\Delta_{vir}(z)\right]^{1/3} (1+z) {\rm keV}\,,
\label{eq:mt}
\ea 
where 76\% of the gas is assumed to be hydrogen (see, e.g., Eke et
al. 1996). $\Delta_{vir}(z)$ is given by the ratio between
the mean density within the virialized region and the average cosmic
density at redshift $z$. For $\Omega_0=1$ it is
$\Delta_{vir}=18\pi^2$, independent of $z$, while we use the
expressions provided by Kitayama \& Suto (1996) for the $\Omega_0=0.3$
cases. For isothermal gas, the $\beta_T$ parameter is defined as the ratio
of the specific kinetic energy of the collisionless matter to the
specific thermal energy of the gas, $\beta_T={\mu m_p\sigma_v^2\over
  k_BT}$, with $\mu=0.59$ the mean molecular weight, $m_p$ the proton
mass and $\sigma_v$ the one--dimensional cluster internal velocity
dispersion.  Equivalence between specific gas
thermal energy and DM kinetic energy implies $\beta_T=1$. However,
neither of such assumptions may be completely correct.  

The calibration of the $\beta_T$ value using numerical simulations
has been attempted by several authors (see, e.g., Bryan \& Norman
1998, for a summary of numerical results).  Recent simulations of the
Santa Barbara Cluster Comparison Project (Frenk et al. 1998), based on
a variety of numerical techniques, indicate that $\beta_T
\simeq 1.15$. This value will be adopted in the following as the
fiducial one, while we will also show the effect of changing this
parameter over a rather broad range. For instance, a smaller $\beta_T$
gives a smaller mass and, therefore, a larger $\sigma_M$,
at a fixed temperature, so as to reduce the biasing factor in
eq.(\ref{eq:beff}). One should however bear in mind that such N--body
calibrations of $\beta_T$ generally include only adiabatic physics of
the intra--cluster medium, while neglecting the effects of radiative
cooling, feedback effects from galactic winds, etc.

As a second step we convert $X$--ray temperatures into bolometric
luminosities, $L_{bol}$. This is probably the most delicate step of our
procedure, since neither observations nor theoretical modelling
converge to a unique, well determined $L_{bol}$--$T$ relation.  By
adopting a phenomenological approach, we model the
$L_{bol}$--$T$ relation as 
\be 
    L_{bol}\,=\,L_6\,\left(T\over 6{\rm
    keV}\right)^\alpha (1+z)^A \times 10^{44} h^{-2}\lum \,\,,
\label{eq:lt}
\ee 
where $L_6$ gives the expected luminosity for a 6 keV cluster, while
$A$ defines the redshift evolution of the $L_{bol}$--$T$ relation. 
Data for local clusters indicates that $L_6\simeq 3$ as a rather
stable result, and $\alpha \simeq 2.5$--3.5, at least for temperatures
$T\magcir 1$ keV, depending on the sample, data analysis technique and
corrections for cooling flow effects.  White, Jones \& Forman (1997)
analysed a set of 207 {\sl EINSTEIN} clusters and found $\alpha \simeq
3$, thus in agreement with previous results (e.g., David et al.  1993,
and references therein). Although the formal fitting uncertainties are
generally small, the scatter of data points around the relation
(\ref{eq:lt}) is so large as to raise the question of whether it
represents a good model for the observational $L_{bol}$--$T$ relation.
A remarkable reduction of the scatter is found once the effects of
cooling flow are properly introduced, at least for temperatures
$k_BT\magcir 4\,$keV (e.g., White et al. 1997; Arnaud \& Evrard 1997;
Markevitch 1998; Allen \& Fabian 1998). In the following we will adopt
the value $\alpha=3$ as the reference one, while we will also show the
effect of its variation 
on the resulting model constraints.  Since available data of the redshift 
dependence of the $L_{bol}$--$T$ relation indicate no evolution up to 
$z\simeq 0.4$ (Mushotzky \& Scharf 1997) we will use $A=0$ in the following.
No dependence of the final results on $A$ are expected, owing to the
limited depth of XBACs.

As a third step we convert bolometric luminosities, $L_{bol}$, into
finite energy--band ([0.1-2.4] keV) luminosities, $L_E$, according to
$L_E=f_E[T(M,z);z]L_{bol}$, where the bolometric correction term,
$f_E$, is computed by integrating the emissivity over the appropriate
energy band.  Following Mathiesen \& Evrard (1998), we assume in our
analysis a purely Bremsstrahlung emissivity, with the power--law
approximation for the Gaunt factor, $g(E,T)\propto (E/kT)^{-0.3}$.
This approximation has been shown to be rather precise for $T_X\magcir
2$ keV, which is the temperature range expected to be relevant for the
bright Abell clusters of XBACs, in correcting luminosities in the soft
{\sl ROSAT} band, [0.5--2.0] keV (Borgani et al. 1998). We expect this
approximation to be at least as good, when dealing with the broader
XBACs band.

\section{Analysis and results}
In order to place constraints on the model parameter space, we should
device a method which is able to estimate the values of the slope
$\Gamma$ of the power spectrum and its amplitude $\sigma_8$, along
with the respective statistical uncertainties, in the most
assumption--free way.

\begin{figure}
%\vspace{5cm}
\centering
\psfig{figure=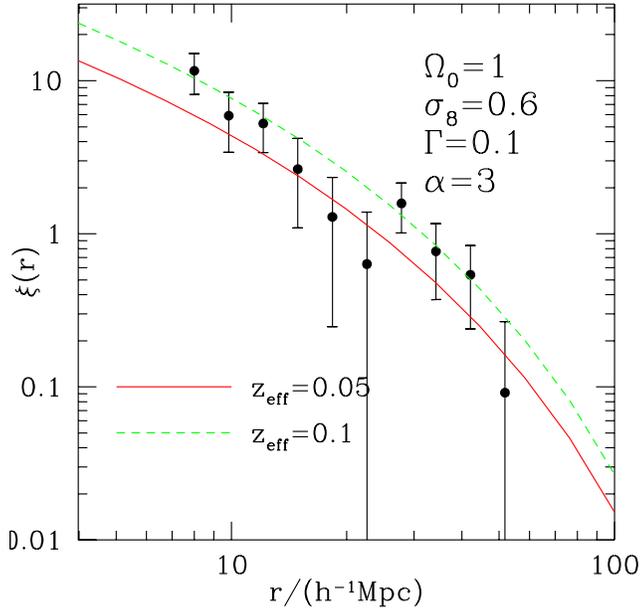,height=8.5cm}
\caption{The influence of the choice for the effective survey
redshift, $z_{eff}$, on the cluster 2--point correlation function
predicted by a specific cosmological model with a fixed $L_X$--$T_X$
relation. The flux--limit corresponds to that of the sample {\sl (a)},
whose, $\xi(r)$ is also plotted as filled circles.
}
\label{fi:xi_zeff}
\end{figure}

Binning the data as in Fig. \ref{fi:xi} may in principle introduce an
uncertainty in the estimated $\xi(r)$, whose value depends in general
on the binning choice. Furthermore, the theoretical $\xi(r)$ of
eq.(\ref{eq:xih}) requires the value of the mass threshold $M$ to be
known. Even for a fixed choice of the mass--luminosity relation, the
value of $M$ depends on the {\em effective} redshift, $z_{eff}$, that
has to be associated to the whole survey. Indeed, for a given
flux--limit, the {\em effective} luminosity limit is given by
\be
L_{eff}\,=\, 4\pi d_L^2(z_{eff}) S_{lim}\,,
\label{eq:leff}
\ee where $d_L$ is the luminosity--distance. In principle, a
reasonable choice for $z_{eff}$ is the peak of $N(z)$, which occurs at
$\sim 0.075 - 0.08$ (see Fig. \ref{fi:Nr}). However, due to the
broadness of the XBACs redshift distribution, the precise value of
$z_{eff}$ can not be fixed in an unambiguous way.  We show in Figure
\ref{fi:xi_zeff} the effect of changing the value of $z_{eff}$ from
0.05 to 0.1 on $\xi(r)$, for a fixed cosmological models and
$L_X$--$T_X$ relation. Note that, for both such values of $z_{eff}$,
the redshift distribution is far from declining and, therefore, any
redshift within this interval can in principle be adopted as
$z_{eff}$. Increasing $z_{eff}$ from 0.05 to 0.1, has a
non--negligible effect on the resulting $\xi(r)$: the larger
$z_{eff}$, the larger the luminosity-- and, therefore, the
mass--threshold. In turn, this corresponds to a larger biasing factor
and, hence, to an increase of $\xi(r)$. Indeed, at $z_{eff}=0.05$, the
mass--threshold is $M_{th}\simeq 2.1\times 10^{14}h^{-1}M_\odot$ which
corresponds to an effective biasing $b_{eff}\simeq 4.6$. Such numbers
increases to $M_{th}\simeq 4.4\times 10^{14}h^{-1}M_\odot$ and 
$b_{eff}\simeq 6.2$ for $z_{eff}=0.1$. 

This demonstrates that any ambiguity in the choice of $z_{eff}$
generates a similar ambiguity in the determination of the
best--fitting model.

Furthermore, at large separations the estimate of $\xi(r)$ is likely
to be dominated by more distant and, therefore, more luminous
clusters. As a consequence, $\xi(r)$ at different scales takes
contributions from cluster populations having different luminosity and,
in principle, different clustering properties. A possible solution to
this problem would be selecting only clusters above a fixed luminosity
limit. It is however clear that this remedy can only be pursued at the
expense of substantially reducing the sample statistics.

To overcome such limitations of the binned $\chi^2_{min}$ approach,
one can estimate the likelihood that the model correlation function
produces the measured number of cluster pairs at a given separation
and at a given redshift, for specified sample flux--limit and
mass--luminosity relation, in a way which does not depend on the
binning procedure.

Let $\xi(r,z,M(z))$ be the model correlation function for halos
that, at redshift $z$, have mass above the value required by the
sample flux--limit. Then, the number of cluster--cluster pairs with
separation between $r-dr/2$ and $r+dr/2$ and at redshift between
$z-dz/2$ and $z+dz/2$ is 
\be
{\cal C}(r,z)\,dr\,dz\,=\,[1+\xi_{cl}(r,z,M(z))]\,{{\cal R}(r,z)\over
2}\,dr\,dz\,.
\label{eq:clcl}
\ee
Here ${\cal R}(r,z)$ is the expected number of cluster--random
pairs, within the same separation and redshift interval. We estimated
${\cal R}(r)$ by averaging over 20,000 random samples. 
The two--dimensional grid in $r$ and $z$ has 50
equal amplitude bins in the separation interval $6\le r \le 56$ \hMpc
and 10 equal amplitude bins for redshift $z\le 0.2$. It is clear that
the bin size can be made as small as desired, by proportionally
increasing the number of random samples, so as to make the final
results independent of the bin size.  The relevant quantity to
estimate is the likelihood function ${\cal L}$, which is defined as
the product of the probabilities of having one pair at each of the
$(r_i,z_i)$ bins occupied by the data cluster--cluster pairs and the
probability of having zero pairs in all the other elements of the
$r$--$z$ plane. By assuming Poisson probabilities for bin occupation,
we get
\ba
{\cal L}(r,z) & = & \prod_i \exp[-{\cal C}(r,z)\,dr\,dz]\,{\cal
C}(r,z)\,dr\,dz \nonumber \\
& \times & \prod_{j\ne i}\exp[-{\cal C}(r,z)\,dr\,dz]\,,
\label{eq:like}
\ea
where the indices $i$ and $j$ runs over the occupied and empty bins,
respectively. The best fitting values of the model parameters
are found by minimising the quantity
$S=-2{\mbox {\rm ln}}{\cal L}$. After dropping all the terms
independent of the cosmological scenario, it reads
\be
S\,=\,2\int dr \,\int dz \,{\cal C}(r,z) -2\sum_i {\mbox {\rm ln}}{\cal
C}(r_i,z_i) \,.
\label{eq:ent}
\ee
A similar maximum--likelihood (ML
hereafter) approach has been applied by Croft et al. (1997) to
estimate the richness dependence of the correlation length $r_0$ of
APM clusters (see also Marshall et al. 1983). 

\begin{figure}
%\vspace{5cm}
\centering
\psfig{figure=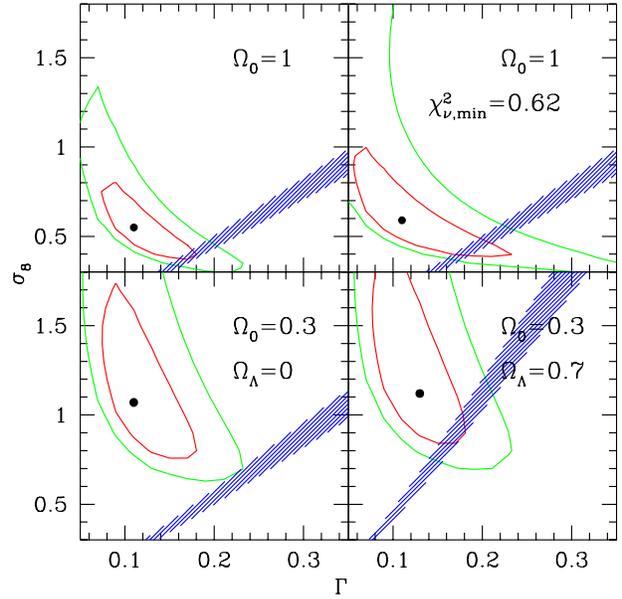,height=8.5cm}
\caption{Iso--$\Delta S$  contours on the $\sigma_8$--$\Gamma$ plane
  from the sample {\sl (a)}, for $\Omega_0=1$ and 0.3, with both flat
  and open geometries. The upper right panel show the results from the
  $\chi^2$--minimisation analysis for the critical--density case.
  Contours correspond to $1\sigma$, $2\sigma$ and $3\sigma$ confidence
  levels for 2 significant fitting parameters.  In each panel, the
  shaded area represents the 1$\sigma$ confidence region from the {\sl
    COBE} normalisation, as given by White \& Scott (1997). }
\label{fi:fitmod}
\end{figure}

The results of the application of the ML analysis are shown in Figure
\ref{fi:fitmod}, for a low--density model with $\Omega_0=0.3$, both
with open and flat geometry, and for the critical density case. While
the selected range of the shape parameter $\Gamma$ turns out to be
almost independent of $\Omega_0$, the value of $\sigma_8$ is a
decreasing function of the density parameter, much like the $\sigma_8$
values coming from the local cluster abundance constraint (e.g., Eke
et al. 1996; Viana \& Liddle 1996; Girardi et al. 1998). It is worth
remembering that the confidence regions in the $\sigma_8$--$\Gamma$
plane from the ML analysis are estimated with the underlying
assumption that cluster pairs are independent of each other. A careful
error estimate would pass through the construction of a non--Gaussian
likelihood function (e.g., Dodelson, Hui \& Jaffe 1997), which in
principle requires the knowledge of the whole $N$--point correlation
function hierarchy. In their ML analysis of APM clusters, Croft et al.
(1998) compared errors from the Poissonian likelihood with the cosmic
variance from N--body simulations. They concluded that accuracy of
errors from a Poissonian ML function depends on the cluster richness,
being rather accurate for cluster populations having a mean separation
$d_{cl}\magcir 70~$ \hMpc.  Assuming $\alpha=3$ in the $L_X$--$T_X$
relation and $\beta_T=1.15$ in the $M$--$T_X$ conversion, the XBACs
flux--limit corresponds to a cluster mass--limit $\simeq 3.8\times
10^{14}h^{-1}M_\odot$ at $z_{eff}=0.08$. In turn, taking the
$\Omega_0=1$ best--fitting model with $\sigma_8=0.55$ and
$\Gamma=0.11$, the resulting cluster abundance from the
Press--Schechter formula would correspond to $d_{cl}\simeq 85$ \hMpc
for the mean separation of clusters above this mass limit. Therefore,
we expect the confidence regions found by the ML approach to provide a
realistic estimate of the actual uncertainties in the parameter
estimates.

In order to check the robustness of the ML results, we also decided to
apply the $\chi^2_{min}$ analysis to the binned $\xi(r)$.
As already mentioned, a potential limitation of the $\chi^2_{min}$
approach lies in the ambiguity in the definition of the effective
redshift of the survey. In this analysis, we decided to assume
$z_{eff}= 0.075$, which corresponds to the peak of the redshift
distribution. The results for the
$\chi^2_{min}$ analysis for $\Omega_0=1$ are reported in the upper
right right panel of Figure \ref{fi:fitmod}. It turns out that the two
methods identify essentially the same regions of the
$\sigma_8$--$\Gamma$ plane. From the one hand, this result supports
the robustness of the analysis methods.  From the other hand, it
indicates that the {\em effective} survey redshift is in this case
reliably represented by the peak of $N(z)$.  We report in Table
2 the best--fitting parameter values, for the three
cosmological models, for both the ML and the $\chi^2_{min}$ method,
along with their 2$\sigma$ uncertainties for one significant
parameter, i.e.  after maginalising with respect to the other
parameter.

We note that $\sigma_8$ takes larger values for the smaller
$\Omega_0$. This is due to the fact that, for fixed cluster distance
and flux--mass conversion, the sample flux limit turns into a larger
value of the cluster Lagrangian radius $R_M$ in eq.(\ref{eq:lr}) when
a smaller density parameter is assumed.  Therefore, for a fixed $P(k)$
amplitude, clusters would correspond to more rare and thus more
clustered peaks. A suitable increase of $\sigma_8$ (i.e., of the
$P(k)$ amplitude) is thus required to decrease the relative height of
the peaks to be associated with clusters.

The resulting constraint on $\sigma_8$ and its dependence on
$\Omega_0$ is comfortably consistent with, although somewhat less
stringent than, that coming from the abundance of local galaxy
clusters. Indeed, several independent analyses based on the optical
virial mass function (e.g., Girardi et al. 1998), the local $X$--ray
temperature function (e.g., Viana \& Liddle 1996; Eke et al. 1996;
Oukbir, Bartlett \& Blanchard 1997) and the local $X$--ray luminosity
function (e.g., Borgani et al. 1998) converge to indicate that
$\sigma_8\simeq 0.5$--0.6 for $\Omega_0=1$, while it rises to
$\sigma_8\simeq 0.9$--1.2 for $\Omega_0=0.3$, almost independent of
the presence of a cosmological constant term.  This shows that
combining results from cluster abundance and clustering do not allow
to break the degeneracy between $\sigma_8$ and $\Omega_0$. In any
case, it is remarkable that both the abundance and the large--scale
distribution of clusters, which involve rather different scale ranges,
can be consistently interpreted in the framework of hierarchical
clustering of Gaussian density fluctuations, as described by the Press
\& Schechter (1974) approach and by its MW extension to the 2--point
correlation function.

As for the shape parameter, it is generally constrained to lie in the
range $0.05\mincir \Gamma \mincir 0.20$, thus in agreement with,
although somewhat smaller than, that obtained from the optical
Abell/ACO cluster distribution (Borgani et al. 1997) and from the APM
galaxy distribution (e.g., Peacock \& Dodds 1994).

The shaded areas in each panel of Fig. \ref{fi:fitmod} represent the
1$\sigma$ constraint from the {\sl COBE} normalisation, as provided by
White \& Bunn (1996). It turns out that, if $\Omega_0=1$,
$\sigma_8\simeq 0.4$--0.5 and $\Gamma \simeq 0.18$ (at 1$\sigma$ c.l.)
are required to satisfy at the same time the large--scale CMB
constraint and the XBACs cluster clustering. As for the low--density
flat model, it requires $\sigma_8\simeq 0.9$--1.1, with essentially
the same shape parameter. As for the open model, CMB and cluster $\xi(r)$ 
constraints are rather inconsistent with each other. 

\begin{figure}
%\vspace{5cm}
\centering
%\mbox{\psfig{figure=xi_like.ps,height=15cm,bbllx=18pt,bblly=144pt,bburx=592pt,bbury=718pt}}
\psfig{figure=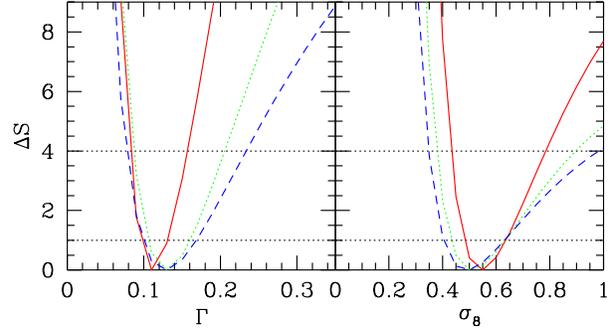,height=8.5cm}
\vspace{-4.truecm}
\caption{The variation of $\Delta S$ around the best--fitting value of
  each parameter, after marginalising with respect to the other
  parameter. Continuous, dotted and dashed curves are for the {\sl
    (a), (b)} and {\sl (c)} samples, respectively. As in Figure
  \ref{fi:fitpow_marg}, the two horizontal lines and the upper limit
  of each panel represent the $1\sigma$, 2$\sigma$ and 3$\sigma$
  confidence levels for one significant parameter.}
\label{fi:marg_fl}
\end{figure}

\begin{figure}
%\vspace{5cm}
\centering
%\mbox{\psfig{figure=xi_like.ps,height=15cm,bbllx=18pt,bblly=144pt,bburx=592pt,bbury=718pt}}
\psfig{figure=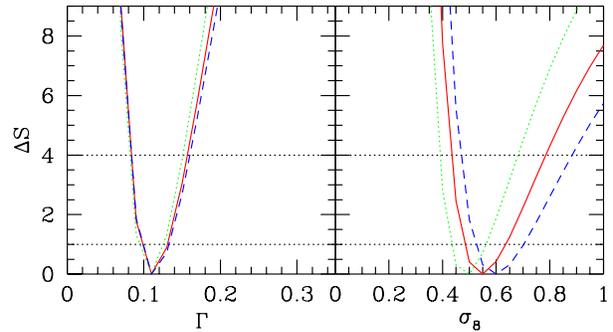,height=8.5cm}
\vspace{-4.truecm}
\caption{The same as Figure \ref{fi:marg_fl}, but for the effect of
  changing the parameter $\alpha$ in the $L_X$--$T_X$ relation.
  Dotted, continuous and dashed curves are for $\alpha=2.5, 3$ and
  3.5, respectively.}
\label{fi:marg_lt}
\end{figure}

\begin{figure}
%\vspace{5cm}
\centering
%\mbox{\psfig{figure=xi_like.ps,height=15cm,bbllx=18pt,bblly=144pt,bburx=592pt,bbury=718pt}}
\psfig{figure=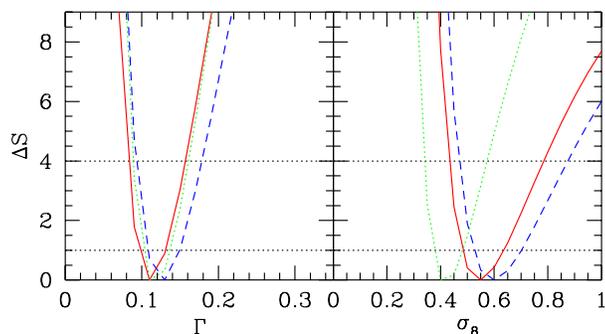,height=8.5cm}
\vspace{-4.truecm}
\caption{The same as Figure \ref{fi:marg_fl}, but for the effect of
  changing the mass--temperature conversion factor, $\beta_T$.  Dotted,
  continuous and dashed curves are for $\beta_T=0.8,1.15$ and 1.5,
  respectively.}
\label{fi:marg_be}
\end{figure}

It is worth reminding here that the above results refer to the whole
XBACs, with $S_{\rm lim}=5 \times 10^{-12}\,$erg s$^{-1}\,$cm$^{-2}$,
as well as to one particular choice of the $L_X$--$T_X$ relation and
of the $M$--$T_X$ conversion factor $\beta_T$.

In Figure \ref{fi:marg_fl} we plot for the $\Omega_0=1$ case the
variation of $\Delta S=S-S_{min}$ around its minimum, as one of the
two parameters $\sigma_8$ and $\Gamma$ is varied, keeping the other
fixed at its best fitting value, for the {\sl (a), (b)} and {\sl (c)}
samples. It turns out that results are quite stable when
varying the flux limit of the XBACs sample analysed; the only significant 
effect being that of enlarging the confidence intervals as the number 
of selected clusters decreases with increasing $S_{lim}$.
The stability of the results against variations of
the flux limit indicates that any possible incompleteness of XBACs at
low fluxes has at most a marginal effect.

The effect of changing the shape of the local $L_X$--$T_X$ relation is
shown in Figure \ref{fi:marg_lt}. The extreme values of $\alpha=2.5$
and 3.5 have been chosen so as to bracket the range of results
presented in the literature (cf. the discussion in the previous
section). Even in this case, constraints on model parameters are
rather robust against changes in the $L_X$--$T_X$ relation. This
represents a remarkable result, since connecting luminosities to
temperatures represents in principle a rather delicate step, owing to
the uncertainties in both observational and theoretical determinations
of the $L_X$--$T_X$ relation.

Finally, we show in Figure \ref{fi:marg_be} the effect of changing the
parameter $\beta_T$ in the $M$--$T_X$ relation. Even in this case, the
extreme values, $\beta_T=0.8$ and 1.5 have been chosen so as to
largely encompass the results from hydrodynamic cluster simulations
(cf. Bryan \& Norman 1998). We find that, while the shape
parameter $\Gamma$ does not depend on $\beta_T$, the value of
$\sigma_8$ systematically increases with $\beta_T$. The reason for
this is that a larger $\beta_T$ implies a larger mass for a fixed
cluster temperature. As a consequence, a larger $\sigma_8$ is required
to suppress the biasing factor, so as to compensate for the selection
of more massive, more clustered clusters.

\begin{table}
\label{t:marg}
\centering
\caption[] {Model parameters for the sample {\sl (a)} from the maximum
likelihood and from the $\chi^2$--minimisation analyses. Uncertainties
corresponds to $2\sigma$ confidence levels for one significant parameter.}
\tabcolsep 4pt
\begin{tabular} {lcccc}
  & \multicolumn{2}{c}{Max-like.} & \multicolumn{2}{c}{$\chi^2$--min.} \\
\noalign{\smallskip}
  & $\sigma_8$ & $\Gamma$ & $\sigma_8$ & $\Gamma$ \\
\noalign{\smallskip}
$\Omega_0=1$ & $0.55^{+0.23}_{-0.11}$ & $0.11^{+0.05}_{-0.03}$ & 
               $0.59^{+0.30}_{-0.12}$ & $0.11^{+0.07}_{-0.04}$ \\
\noalign{\smallskip}
$\Omega_0=0.3;\Omega_\Lambda=0$ & $1.07^{+0.88}_{-0.27}$ & $0.11^{+0.06}_{-0.04}$ & 
                 $1.28^{+0.96}_{-0.26}$ & $0.11^{+0.12}_{-0.06}$ \\
\noalign{\smallskip}
$\Omega_0=0.3;\Omega_\Lambda=0.7$ & $1.12^{+0.88}_{-0.29}$ & $0.13^{+0.06}_{-0.05}$ & 
                 $1.50^{+1.33}_{-0.47}$ & $0.11^{+0.13}_{-0.06}$ 
 
\end{tabular}
\end{table}

\section{Conclusions}
We presented an analysis of the redshift--space 2--point correlation
function for the X--ray Brightest Abell--type Cluster sample (XBACs;
Ebeling et al. 1996), with the aim of investigating the resulting
constraints on cosmological models.  In our analysis, cosmological
models are specified by three parameters, namely the density parameter
$\Omega_0$ (with or without a cosmological constant term to provide
spatial flatness), the fluctuation power--spectrum amplitude through
the $\sigma_8$ quantity and the shape $\Gamma$ of the power--spectrum.

As a starting point, we follow the analytical approach by Mo \& White
(1996), which provides the 2--point correlation function, $\xi(r)$,
for the distribution of virialized halos above a given mass limit
(e.g., Mo, Jing \& White 1996). In order to convert masses into
observed fluxes, we followed a purely empirical recipe (see also
Borgani et al. 1998), which depends on two parameters, namely the
conversion factor $\beta_T$ to pass from mass to $X$--ray temperatures
[cf. eq.(\ref{eq:mt})] and the slope $\alpha$ of the local
$L_X$--$T_X$ relation [cf. eq.(\ref{eq:lt})].

Although this analysis is limited to nearby ($z\mincir 0.2$) clusters,
the formalism that we introduced can be directly applied to any survey
of high--redshift objects. 

The comparison between model and data has been performed by resorting
both to a $\chi^2$--minimisation ($\chi^2_{min}$) procedure and to a
maximum likelihood (ML) approach, which avoids any ambiguity
associated to the binning procedure required by the $\chi^2_{min}$
method.

The main results of our analysis can be summarised as follows.
\begin{description}
\item[(a)] If the 2--point correlation function is modelled as a
  power--law, $\xi(r)=(r_0/r)^\gamma$, then the best fitting
  parameters for the whole sample are $r_0=26.0\pm 4.5$ \hMpc and
  $\gamma=2.0\pm 0.4$. The clustering strength increases very weakly
  as a higher flux--limit is imposed and higher luminosity clusters
  are selected.  As long as a correlation exists between cluster
  luminosity and richness, this result is consistent with the picture
  of a mild dependence of the correlation amplitude on the cluster
  richness (cf. Croft et al. 1997).
\item[(b)] For our reference choice for the mass--luminosity
  conversion (i.e., $\alpha=3$ and $\beta_T=1.15$), we find that
  $\sigma_8 \simeq 0.6\pm 0.2$ (2$\sigma$ uncertainties after
  marginalisation), while it becomes about twice as large for
  $\Omega_0 =0.3$. This result is completely consistent with
  constraints coming from cluster abundances and indicates that the
  picture of hierarchical clustering of Gaussian primordial density
  fluctuations is able to account at the same time for both the
  abundance and the clustering of galaxy clusters.
  
  As for the shape of $P(k)$ it is much less dependent on $\Omega_0$,
  with values ranging in the interval $0.05\mincir \Gamma \mincir
  0.20$. Such results are left unchanged by increasing the sample
  flux--limit.
\item[(c)] Adding also the large--scale CMB constraints, we find that an
  $\Omega_0=1$ model with $\sigma_8\simeq 0.4$--0.5 and $\Gamma\simeq
  0.16$--0.20 and a flat $\Omega_0=0.3$ model with $\sigma_8\simeq
  0.9$--1.1 and the same $\Gamma$ are consistent with both the {\sl
    COBE} data and the clustering of XBACs clusters. For the open
  $\Omega_0=0.3$ model, the {\sl COBE} and XBACs constraints are quite
  inconsistent with each other.
\item[(d)] As for the robustness of such results against changes of
  the mass--luminosity relation, it turns out that $\Gamma$ is quite 
  insensitive to both the mass--temperature conversion
  factor $\beta_T$ and to the slope $\alpha$ of the local $L_X$--$T_X$
  relation. The power spectrum amplitude $\sigma_8$ has only a weak 
  dependence on the choice for the
  $L_X$--$T_X$ relation, while it increases from about 0.35 to 0.6 as
  $\beta_T$ varies from 0.8 to 1.5.
\end{description}

As a concluding remark, we point out that, although XBACs provides
useful constraints on cosmological models, it is not a completely
$X$--ray selected sample. Newer and larger $X$--ray cluster samples
have been recently compiled (BCS; Ebeling et al. 1998) or are close to
completion (REFLEX; B\"ohringer et al. 1995). Such samples, thanks to
their careful definition of completeness criteria and to their
increased statistics will allow to definitely fix the clustering
scenario for local $X$--ray clusters. Even more exciting, the
possibility of realizing cluster surveys at higher redshifts with
next--generation $X$--ray satellites will provide a further means to
constrain both the power spectrum of density fluctuations and the
value of the matter density parameter.

\section*{Acknowledgements}
We wish to thank Marisa Girardi, Luigi Guzzo, Sabino Matarrese,
Lauro Moscardini and Piero Rosati for useful discussions.

\end{document}